\documentclass[prb,twocolumn,amssymb,superscriptaddress]{revtex4-2}
\usepackage{amsmath}
\usepackage{amssymb}
\usepackage{amsthm}
\usepackage{amsfonts}
\usepackage[version=4]{mhchem}
\usepackage{listings}
\usepackage{enumerate}
\usepackage{latexsym}
\usepackage{psfrag}
\usepackage{bm}
\usepackage[all]{xy}
\usepackage{graphicx}
\usepackage[pdftex,colorlinks]{hyperref}
\usepackage{color}
\usepackage{times}
\usepackage{array}
\usepackage{placeins}

\begin{document}

\title{Lattice dynamics  in the charge-density-wave metal at a van-Hove-singularity filling }
\author{Jia-Wei Mei}
\email{meijw@sustech.edu.cn}
\author{Fei Ye}
\email{yef@sustech.edu.cn}
\affiliation{Shenzhen Institute for Quantum Science and Engineering, and
   Department of Physics, Southern University of Science and Technology,
   Shenzhen 518055, China}
\affiliation{Shenzhen Key Laboratory of Advanced Quantum Functional Materials
   and Devices, Southern University of Science and Technology, Shenzhen 518055, China}
\author{Xiaobin Chen}
\email{chenxiaobin@hit.edu.cn}
\affiliation{School of Science and State Key Laboratory on Tunable Laser Technology and Ministry of Industry and Information Technology Key Lab of
Micro-Nano Optoelectronic Information System, Harbin Institute of Technology, Shenzhen 518055, China}
\date{\today}

\begin{abstract}
The charge-density-wave (CDW) order with macroscopically occupied electrons
distorts the underlying lattice and usually causes the softening of the
associated phonon mode. However, previous studies demonstrated that the
spin-Peierls transition does not always induce an associated phonon softening,
but the central-peak scenario applied in the quasi-one-dimensional compound
CuGeO$_3$. We generalize the lattice-dynamics studies on the two-dimensional CDW
state at van-Hove-singularity (VHS) filling and find that the CDW ordering could
develop a central peak at zero frequency while the associated phonon undergoes
hardening. The particle-hole scatterings between VHS points give rise to a
low-energy increased charge-density susceptibility, and their coupling to the
lattice dynamics induces two poles in the Green function for the CDW-associated
phonon mode. The zero-frequency pole corresponds to the collective
charge-density and phonon coupling mode. The high-frequency one is related to
the high-temperature phonon mode that hardens as reaching the CDW transition.
Our result may have the potential implication for the recently discovered Kagome metal $A$V$_3$Sb$_5$ ($A$ = K, Rb, Cs) in which no soft phonon is observed during the CDW transition.  
\end{abstract}

\maketitle

\emph{Introduction.--}
The low-energy excitation in metal gives rise to an increased bare electron charge-density susceptibility $\chi_0(\mathbf{Q})$ with the wave vector $\mathbf{Q}$ connecting two extreme points at the Fermi surface. Thus a periodic potential, e.g. through the electron-phonon interaction, will cause an electronic charge-density modulation as $\Delta\rho(\mathbf{Q})=\chi^0(\mathbf{Q})V(\mathbf{Q})$. As feedback, the charge-density fluctuation modifies the periodic potential. It renormalizes the phonon frequency, $\omega^2(\mathbf{Q})=\Omega_0^2(1+\kappa\chi_0(\mathbf{Q}))$, where $\Omega_0$ is the bare phonon spectrum, leading to the softening of $\omega(\mathbf{Q})$, known as the Kohn anomaly~\cite{Kohn1959}. When it reduces the phonon frequency to zero, the charge-density modulation induces a static lattice distortion, leading to the charge-density-wave (CDW) order~\cite{Gruener1988, Monceau2012}. Rare-earth tritellurides $R$Te$_3$ are prime examples of CDW ordered ground states where the phonon softening appears near the CDW transition temperature $T_{\rm CDW}$~\cite{Ru2008, Hoesch2009, Maschek2018}. Unexpectedly, in the recent kagome compounds $A$V$_3$Sb$_5$ ($A$ = K, Rb, Cs), which exhibit intertwined CDW order and superconductivity~\cite{Ortiz2019,Ortiz2020,Ortiz2021,Ortiz2021a}, no CDW-associated phonon softening is observed in the X-ray and neutron scatterings~\cite{Li2021, Xie2021}. Even more puzzling, the neutron scattering reveals the phonon hardening during the CDW transition in CsV$_3$Sb$_5$~\cite{Xie2021}. The origin of the CDW in $A$V$_3$Sb$_5$ probably mainly arises from collective particle-hole scattering between the van-Hove-singularity (VHS) $M$-points~\cite{Feng2021,Park2021,Wu2021,Liu2021a}, which enhances the charge-density fluctuations and modifies the lattice dynamics.

The puzzled phonon behaviors in $A$V$_3$Sb$_5$ motivate us to study the lattice
dynamics in the CDW metal at the kagome VHS filling in this paper. To simplify the problem,
we study a phenomenological model that describes the coupling of a single
CDW-associated phonon mode and the charge-density fluctuations near the Fermi
surface on the random-phase-approximation (RPA) level. Such a model belongs to
the Fano problem for a discrete phonon mode embedding in the 
charge-density-fluctuation continuum. We tune the electron-electron interaction $u$ and
electron-phonon coupling $\kappa$ to induce the CDW transition within RPA 
phenomenologically. Within proper parameters in the CDW metal at VSH filling, a
central peak appears in the phonon dynamic structural factor as we approach the CDW
transition. Meanwhile, the CDW-associated phonon frequency becomes hardening, instead of
softening.  The related central-peak physics has previously been discussed similarly in
the famous spin-Peierls insulating compound
CuGeO$_3$~\cite{Gros1998,Ye2001,Holicki2001}, in which the phonon frequency is
much larger than the spin-Peierls transition temperature. For a comparative
study, the soft phonon mode is found in the lattice dynamics in the CDW order induced by the
nesting of nearly parallel quasi-1D Fermi surface, consistent with experiments
in the CDW compound $R$Te$_3$. By analysis of the pole  structure,  
the phonon Green function has one pole for the soft-phonon
scenario while it has two poles in the central-peak regime in which the
zero-frequency pole corresponds to the collective charge-density and phonon
coupling mode and the high-frequency one is related to the high-temperature
phonon mode. Our simple RPA theory may have the potential implication for the absence of
the soft-phonon mode in $A$V$_3$Sb$_3$ although its more complicated multiple-band structure crossing the Fermi level requires further investigations.

\begin{figure*}[t]
  \centering
  \includegraphics[width=2\columnwidth]{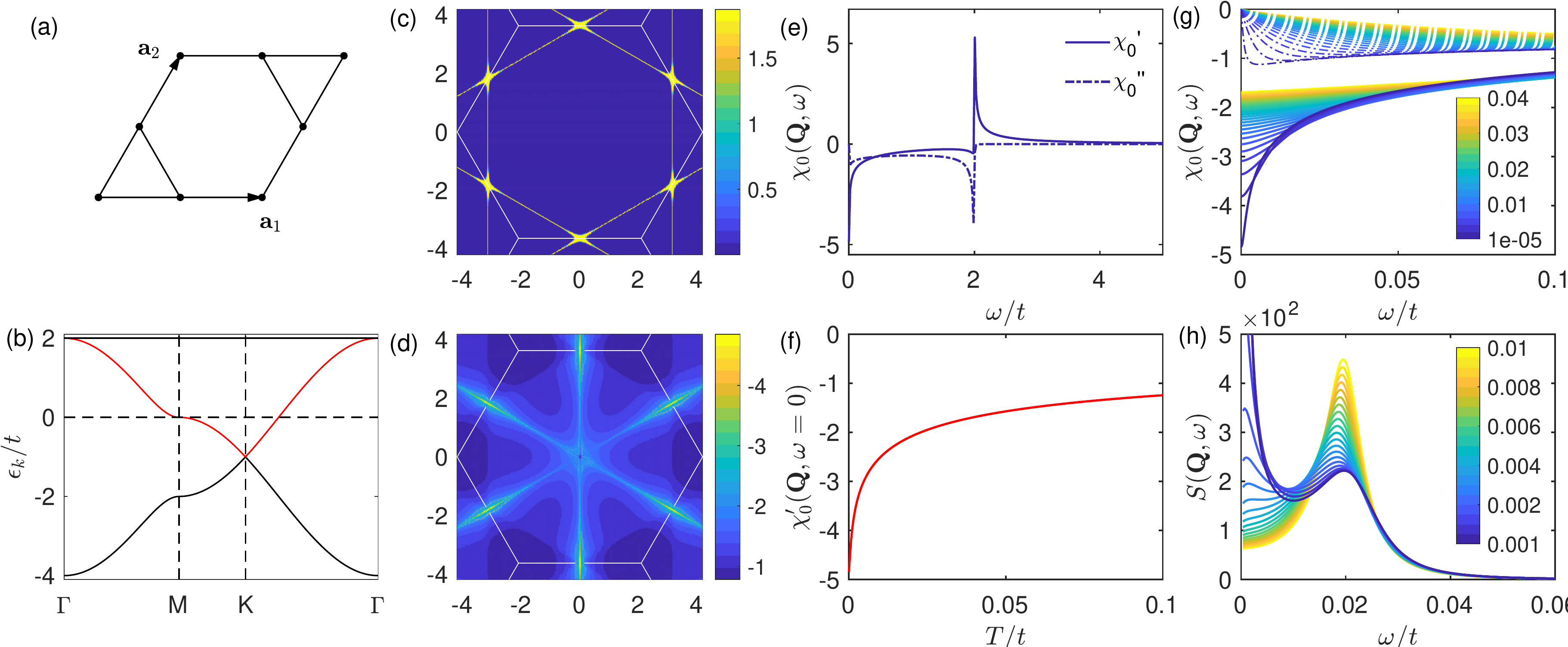}
  \caption{Central peak for kagome VHS filling. (a) Kagome lattice with primitive unit vectors $a_1=(1,0)$ and
    $a_2=(1/2,\sqrt{3}/2)$. (b) Band dispersion of the simple kagome model
    along the high-symmetry direction. (c) $\mathbf{k}$-dependent 
    ${\rm DOS}(\mathbf{k})$ as in Eq.~(\ref{eq:dos}) where the VHS points at $M$
    displays the enhanced value. (d) $\mathbf{q}$-dependent static
    charge-density susceptibility $\chi_0(\mathbf{q})$ at low temperature
    $T=0.00001t$ with an enhanced amplitude at the CDW wave vectors $\mathbf{Q}$
    with $(k_1,k_2)=(0,\pi),(\pi,0),(\pi,\pi)$. (e) $\omega$-dependent
    $\chi(\mathbf{Q},\omega)$ at low temperature $T=0.00001t$. The solid and
    dotted lines are the real and 
    imaginary parts, respectively. (f) Temperature-dependent static charge-density susceptibility 
    $\chi(\mathbf{Q},\omega=0)$. (g) Temperature evolution of $\omega$-dependent 
    $\chi(\mathbf{Q},\omega)$.  The solid and dotted lines are the real and
    imaginary parts, respectively. (h) Temperature evolution of the dynamical
    structure factor $S(\mathbf{Q},\omega)$ with $u/\kappa=0$ for $T_{\rm CDW}=0.001t$. The color bars in (g) and (h)
    represent the temperature value in the unit of $t$.}
  \label{fig:figure1}
\end{figure*}

\emph{RPA methods and related experimental parameters. --}
Given the band dispersion $\epsilon_{\mathbf{k}}$, the
density of states (DOS) at Fermi level with $\omega=0$ is approximated as
\begin{eqnarray}
  \label{eq:dos}
  {\rm DOS}(\mathbf{k})=-\frac{2}{\pi}{\rm
  Im}\int_{-\Delta\omega}^{\Delta\omega}\frac{d\omega'}{\omega'+i\delta-\epsilon_{\mathbf{k}}},
\end{eqnarray}  
and the charge density $\rho(\mathbf{q})=\sum_{\mathbf{k}\sigma}c_{\mathbf{k}+\mathbf{q}\sigma }^\dag
c_{\mathbf{k}\sigma }$ has the susceptibility  
\begin{eqnarray}
  \label{eq:P} 
  \chi(\mathbf{q},i\omega_n)&=&\frac{1}{V}\int_0^\beta e^{i\omega_n\tau}\rho(\mathbf{q},\tau)\rho(-\mathbf{q},0)d\tau,
\end{eqnarray}
which represents the particle-hole bubble diagram for non-interacting electrons
\begin{eqnarray}
  \label{eq:Pab} 
  \chi_0(\mathbf{q},i\omega_n)
                              =2\int\frac{d^2k}{(2\pi)^2}\frac{n_F(\epsilon_{\mathbf{k}})-n_F(\epsilon_{\mathbf{k}+\mathbf{q}})}{i\omega_n+\epsilon_{\mathbf{k}}-\epsilon_{\mathbf{k}+\mathbf{q}}}.
\end{eqnarray}
On the RPA level, the electron-electron interaction renormalizes the susceptibility
\begin{eqnarray}
  \label{eq:rpaP}
  \chi(\mathbf{q},i\omega_n)=\frac{\chi_0(\mathbf{q},i\omega_n)}{1-u\chi_0(\mathbf{q},i\omega_n)}.
\end{eqnarray}
The effective interaction, e.g. the off-site interactions~\cite{Wang2013}, causes the CDW transition at the CDW wave vector $\mathbf{Q}$ if
\begin{eqnarray}
  \label{eq:cdw_u}
  1-u\chi_0(\mathbf{Q},0)=0.
\end{eqnarray}
The collective charge-density mode  leads to the RPA phonon Green function
\begin{eqnarray}
  \label{eq:rpaPh}
  D(\mathbf{Q},i\omega_n)=\frac{2\Omega_{0}}{(i\omega_n)^2-\Omega_{0}^2(1+\kappa \chi(\mathbf{Q},i\omega_n))},
\end{eqnarray}
where $\Omega_0$ is the bare frequency and $\kappa$ denotes the
electron-phonon coupling strength. The macroscopically occupied CDW electrons
will distort the lattice when
\begin{eqnarray}
  \label{eq:cdw_k}
  1+\kappa\chi(\mathbf{Q},0)=0.
\end{eqnarray}
The phonon dynamic
structure factor is obtained by taking the analytical
continuation $i\omega_n\rightarrow \omega+i\delta$
\begin{eqnarray}
  \label{eq:dsf}
  S(\mathbf{q},\omega)=\frac{-{\rm Im}{D(\mathbf{q},\omega+i\delta)}}{1-e^{-\beta\omega}}.
\end{eqnarray}

The typical energy for the CDW metals $A$V$_3$Sb$_5$ and $R$Te$_3$ has the order
of 1~eV~\cite{Ortiz2020,Ru2008}. During the numerical calculations, we take the hopping amplitude $t$ as
the energy unit that has the typical value of $t=1$~eV.  The
CDW-activated modes in  $A$V$_3$Sb$_5$ and $R$Te$_3$ has the typical bare
frequency $\Omega_0=0.03t$~\cite{Ru2008,Xie2021}. Once we set the CDW transition
temperature as $T_{\rm CDW}$, the interaction parameters $u$ and $\kappa$ are
specified phenomenologically according to the conditions in Eqs.(\ref{eq:cdw_u})
and (\ref{eq:cdw_k}). We can tune the electron-electron and electron-phonon
interactions ratio $u/\kappa$ to study the evolution of the lattice dynamics. The low-temperature charge susceptibility
is sensitive to the Fermi surface structure, and we firstly set $T_{\rm CDW}=0.001t$
and $u/\kappa=0$ to study the phonon dynamic structure factor. The experimental
typical CDW transition temperature in $A$V$_3$Sb$_5$ and $R$Te$_3$ is $T_{\rm
  CDW}=0.01t$, the same order as $\Omega_0$. We then set the transition
temperature as $T_{\rm CDW}=0.01t$ and tune the ratio $u/\kappa$ to discuss the
relevant lattice dynamics. The DOS at Fermi surface is evaluated with the
integration window $\Delta\omega=0.01t$ in Eq.(\ref{eq:dos}) and the smearing
energy takes the value of $\delta=0.001t$ for the analytical continuation $i\omega_n\rightarrow\omega+i\delta$.

\emph{Central peak and phonon hardening for kagome VHS filling. --}
The simple kagome band has the dispersion
$\epsilon_{\mathbf{k}1,2}=(\mp\sqrt{3+2(\cos(k_1)+\cos(k_2)+\cos(k_1-k_2))}-1)t$
and $\epsilon_{\mathbf{k}3}=2t$. Here $k_{1,2}=\mathbf{k}\cdot\mathbf{a}_{1,2}$.
$\mathbf{k}$ and $\mathbf{a}_{1,2}$ are the wave vector and primitive unit
vectors, respectively, of kagome lattice (Fig.~\ref{fig:figure1}(a)).
Fig.~\ref{fig:figure1}(b) is the band dispersion along the high-symmetry
direction in the Brillouin zone with the 2nd band $\epsilon_{\mathbf{k}2}$,
marked in red, crossing the Fermi level. Fig.~\ref{fig:figure1}(c) is the DOS at
Fermi level in which the Fermi surface is composed of corner-shared triangles
with the increased VHS DOS at $M$ points.
The 2nd band dispersion $\epsilon_{\mathbf{k}2}$ determines the low-energy properties of the kagome system and the  charge density is $\rho(\mathbf{q})=\sum_{\mathbf{k}\sigma}c_{\mathbf{k}+\mathbf{q}\sigma 2}^\dag c_{\mathbf{k}\sigma 2}$. For the sake of simplicity, we ignore the local-site overlap matrix element $\sum_{\mathbf{k}}u_{l}^*(\mathbf{k})u_{l'}(\mathbf{k}+\mathbf{q})$ with the local electron Bloch wave package $u_{l}(\mathbf{k})$~\cite{Yu2012,Kiesel2012,Wang2013}. The electron-phonon interaction matrix element is band-dependent and somehow includes the charge-density matrix element. Our results also apply to the simple square and triangular lattice at the VHS filling where the matrix element is constant. 

We numerically evaluate the charge-density susceptibility $\chi_0(\mathbf{q})$
at low temperature $T=0.00001t$ over the entire Brillouin zone as shown in
Fig.~\ref{fig:figure1}~(d). The VHS at $M$ points favors the particle-hole
scatterings connected by the CDW vectors $\mathbf{Q}$ with
$(k_1,k_2)=(0,\pi),(\pi,0),(\pi,\pi)$, equivalent to $M$ vectors in the
Brillouin zone. Such scattering gives rise to the largest amplitude $\chi'_0(\mathbf{Q},0)$
with a logarithmic singularity. The branch cuts of $\chi_0(\mathbf{q})$ along
$\Gamma$-$M$ have the extremes due to the nesting of Fermi surfaces away from VHS. 
\begin{figure}[b]
  \centering
  \includegraphics[width=0.7\columnwidth]{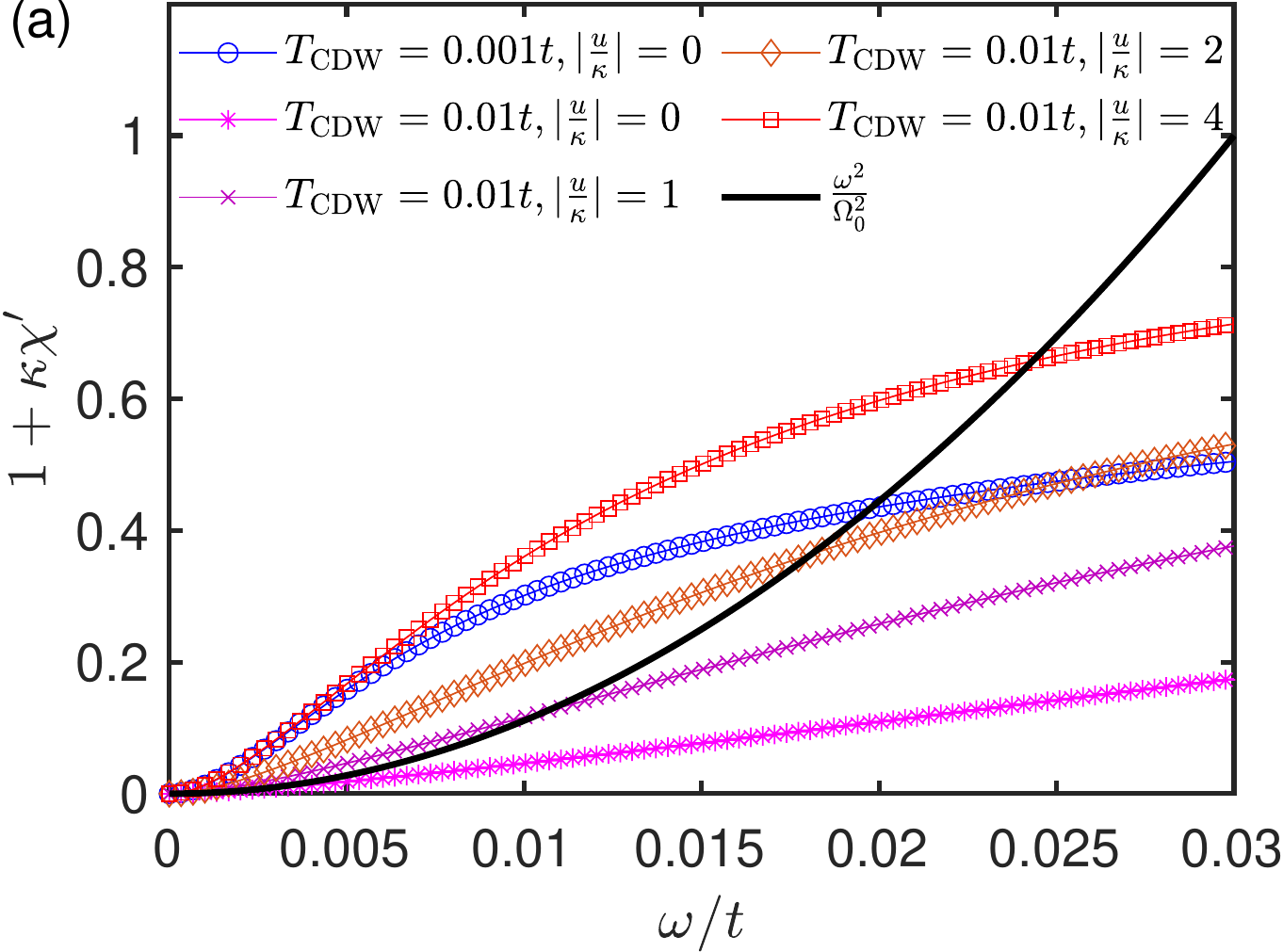}
  \includegraphics[width=\columnwidth]{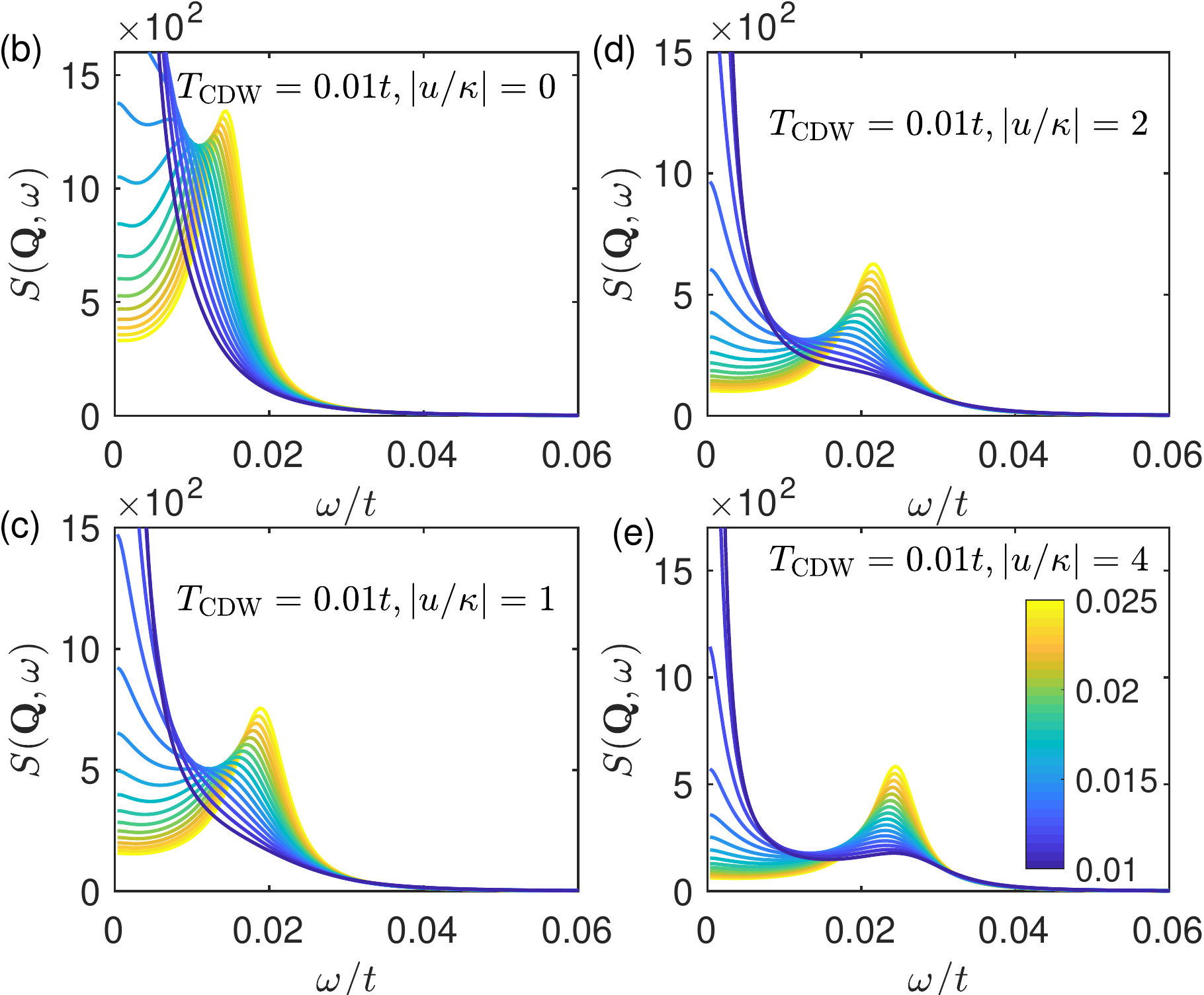}  
  \caption{Central peak for kagome VHS filling with different interaction parameters. (a) Pole analysis of the
    retarded phonon Green function. Temperature evolution of the dynamical
    structure factor $S(\mathbf{Q},\omega)$ with  $T_{\rm CDW}=0.01t$ with
    different interaction parameters with $|u/\kappa|=0$ in (b), 
    $|u/\kappa|=1$ in (c), $|u/\kappa|=2$ in (d), and $|u/\kappa|=4$ in (e). }
  \label{fig:figure3}
\end{figure}
Fig.~\ref{fig:figure1} (e), (f) and (g) display the charge density susceptibility at the CDW wave
vectors $\mathbf{Q}$ at low temperatures
 and low frequencies. The scattering between VHS points results in a logarithmic
divergence in the real part of $\chi'_0(\mathbf{Q})$ at low frequencies
($\omega<0.01t$) and temperatures ($T<0.01t$).  

We set $T_{\rm CDW}=0.001t$ to
see the lattice dynamics when the phonon is coupled to the VHS fluctuations at
low temperature. We
set $u=0$ and fine tune $\kappa$ to make sure $T_{\rm CDW}=0.001t$  and evaluate the phonon dynamical structure
factor 
as shown in Fig.~\ref{fig:figure1}(h). 
The lattice dynamics is a Fano problem where the discrete phonon mode interacts
with the charge-density-fluctuation continuum. The charge-density fluctuation renormalizes the phonon dynamic structural factor, resulting in an asymmetric Fano line shape at high temperatures. With decreasing temperature, the charge-density fluctuation increases at low frequency due to scattering between VHS points and develops into a low-frequency central peak in the phonon dynamic structure factor, in contrast to the Kohn Anomaly in the usual CDW metal. Furthermore, the phonon mode does not show softening as approaching $T_{\rm CDW}$ but hardens at low temperatures.

The above phonon properties are dubbed as the central-peak scenario and applied to the 1D spin-Peierls compound CuGeO$_3$~\cite{Gros1998,Ye2001,Holicki2001}. The phonon Green function has two poles in the central-peak regime, in contrast to a single pole in the soft phonon scenario.  
We analyze the pole of the retarded phonon Green function determined by the equation
\begin{eqnarray}
  \label{eq:pole}
  \frac{\omega^2}{\Omega_{0}^2}-(1+\kappa\chi'(q,\omega))=0,
\end{eqnarray}
by plotting lines for the functions $\frac{\omega^2}{\Omega_{0}^2}$ and
$1+\kappa\chi'(q,\omega)$ in Fig.~\ref{fig:figure3}~(a). For $T_{\rm CDW}=0.001t$ with $|\frac{u}{\kappa}|=0$,
besides the intentional pole at zero frequency, the phonon Green function has
another pole at $\omega\sim0.02t$, related to the bare phonon frequency
$\Omega_0=0.03t$. The two poles lead to the central peak at zero frequency and
the phonon hardening for the lattice dynamics that is coupled to the
charge-density dynamics. 

We set $T_{\rm CDW}=0.01t$ and vary the parameters $u$ and $\kappa$ to study the
lattice dynamics. From Fig.~\ref{fig:figure3}~(a), the phonon Green function has
the single-pole at zero frequency for $u/\kappa=0$. As expected, the lattice
dynamics exhibit the mode softening with decreasing temperature as shown in
Fig.~\ref{fig:figure3}~(b). As we increase the ratio,  the phonon Green function
for $|u/\kappa|=1$ has an extra pole at finite frequency, implying the
central-peak scenario. Two poles are close to each other and different to be
resolved in the dynamic structure factor as shown in Fig.~\ref{fig:figure3}~(c).
With further increasing $|u/\kappa|$, the extra pole appears at a higher
frequency.  The central peak at zero frequency and the high-frequency phonon
hardening are resolved in Fig.~\ref{fig:figure3}~(d) and (e). 

The significant low-energy dependence of the charge-density susceptibility
guarantees the two-pole structure in the phonon Green function when
$1+\kappa\chi'(\mathbf{Q},\omega)$ increases fast than the parabolic function
$\frac{\omega^2}{\Omega_0^2}$. In CuGeO$_3$, the associated CDW phonon frequency
is ten times larger than the spin-Peierls transition temperature. Thus the
two-pole structure is easily achieved. In $A$V$_3$Sb$_5$, the  CDW-related
phonon has a comparable energy scale as the transition temperature. We need a
modest effective electron-electron interaction to increase the low-energy
dependence of $\chi'(\mathbf{Q},\omega)$ for the central peak and phonon
hardening. A VHS-increased low-energy charge-density susceptibility and a modest
effective electron-electron interaction are essential for the central-peak
regime.

\emph{Phonon softening for nearly parallel Fermi surface nesting. --}
\begin{figure}[b]
  \centering
  \includegraphics[width=\columnwidth]{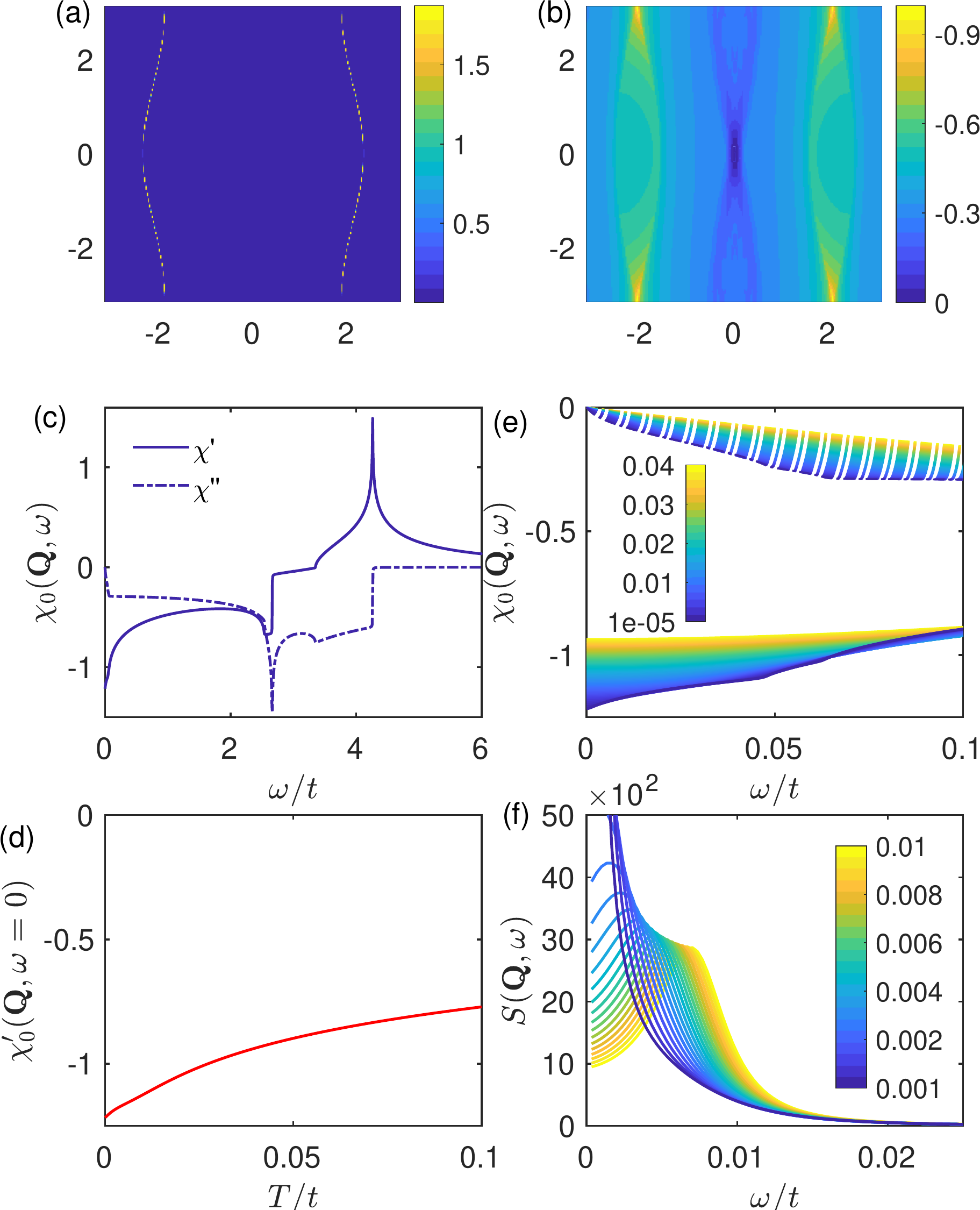}
  \caption{Phonon softening for nearly parallel Fermi surface nesting. (a)-(f) correspond to (c)-(h) in Fig.~\ref{fig:figure1}.}
  \label{fig:figure4}
\end{figure}
\begin{figure}[t]
  \centering
  \includegraphics[width=0.7\columnwidth]{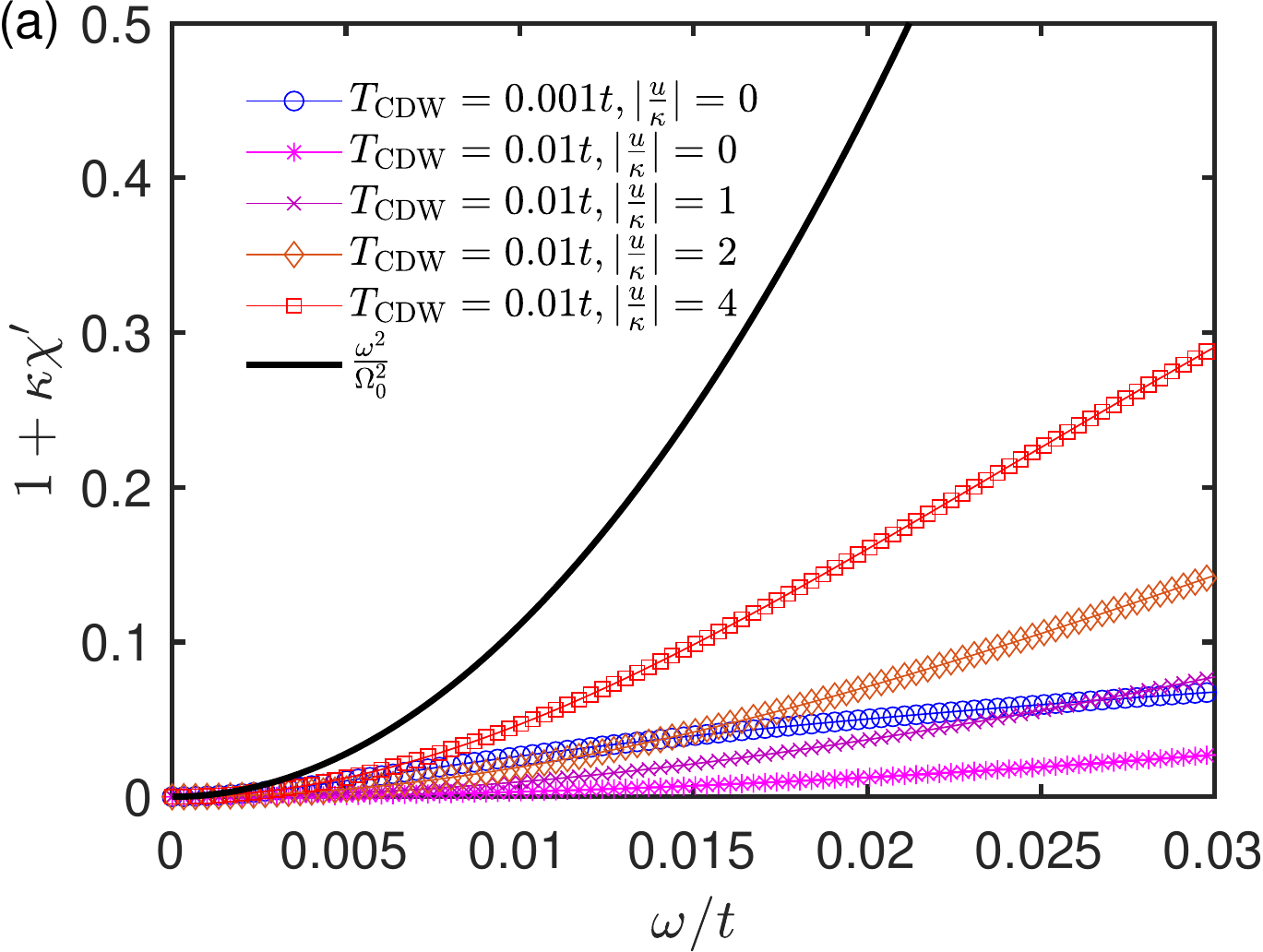}
  \includegraphics[width=\columnwidth]{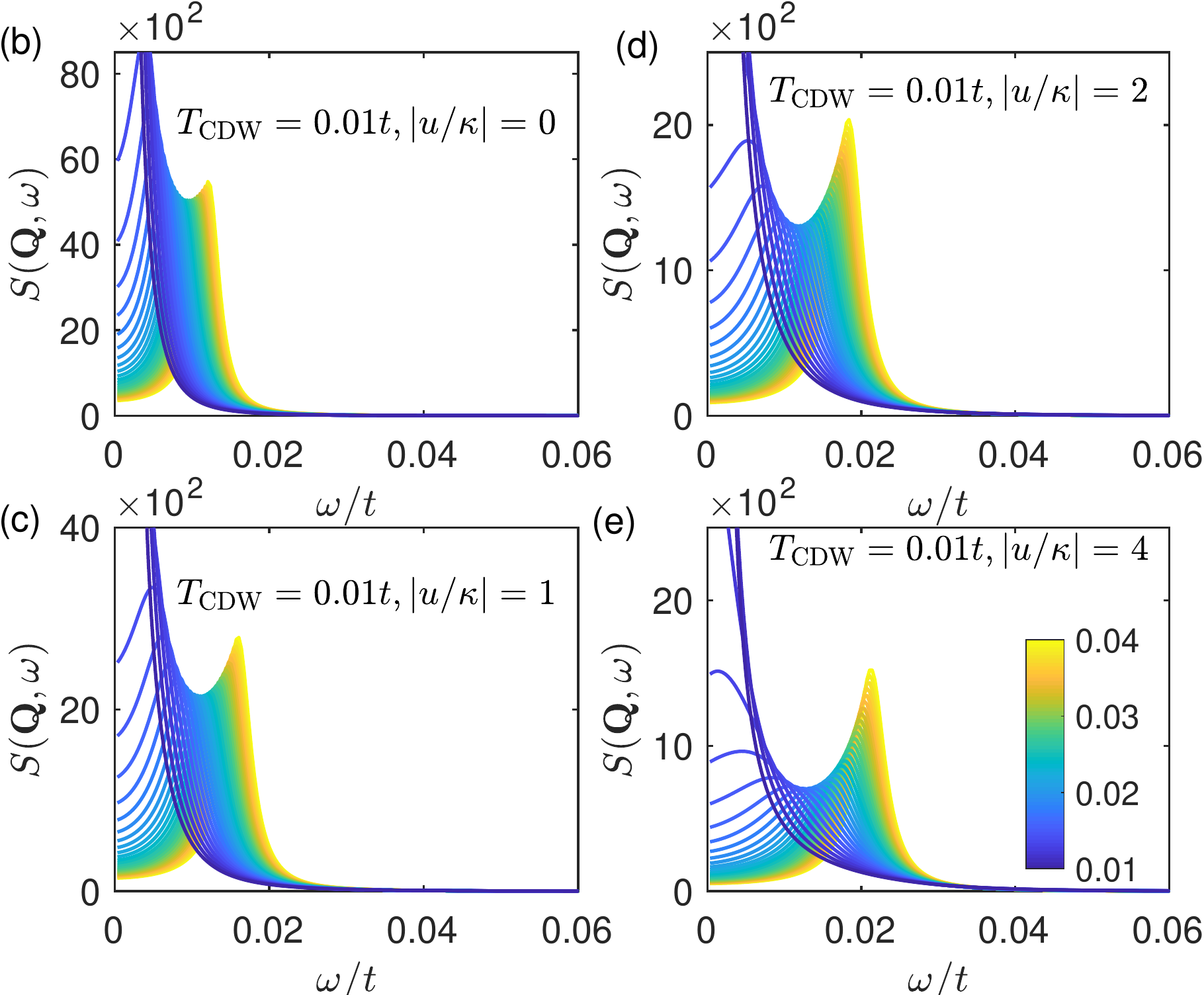}
  \caption{Results of the CDW-activated phonon softening for nearly parallel Fermi surface
    nesting, similar to Fig.~\ref{fig:figure3}. }
  \label{fig:figure6}
\end{figure}
Rare-earth tritellurides $R$Te$_3$ have nearly quasi-1D Fermi surface and its
underlying CDW mechanism is similar to the spin-Peierls transition in CuGeO$_3$.
$R$Te$_3$ has the nearly parallel Fermi surface for the quasi-1D dispersion
$\epsilon_{\mathbf{k}}=-2t(\cos(k_x)+\Delta\cos(k_y))-\epsilon_F$~\cite{Ru2008,Yao2006},
and we set $\Delta=0.2$ and $\epsilon_F=t$. Figs.~\ref{fig:figure4}(a)-(f) are
results corresponding to those for the kagome VHS filling in
Figs.~\ref{fig:figure1}(c)-(h).

With $\Delta=0$, the Fermi surface in Fig.~\ref{fig:figure4}~(a) is the 1D system with two parallel planes where the Fermi vector is $k_F=\pm\frac{\pi}{3}$ and the nesting vector is $Q_x=2k_F$. The finite $\Delta$ induces the $k_y$ dispersion, and the Fermi surface has a nesting with the wave vector $\mathbf{Q}=(2k_F,\pi)$. 
Compared to the VHS filling results in Fig.~\ref{fig:figure1}, the charge-density susceptibility $\chi_0(\mathbf{q})$ in Fig.~\ref{fig:figure4} distributes smoother in the Brillouin zone, consistent with  multiple competing CDW wave vectors $\mathbf{Q}$ in $R$Te$_3$~\cite{Ru2008,Yao2006}. Although there are high-energy edge singularities, the low-frequency $\chi_0(\mathbf{q},\omega)$ is quite smooth in the frequency- and temperature-dependence. By setting $T_{\rm CDW}=0.001t$ with  $u/\kappa=0$, the dynamic structure factor in Fig.~\ref{fig:figure4}(f) displays the phonon softening and becomes divergent as decreasing the temperature to the CDW transition temperature. 

The phonon softening is also resolved in the pole analysis of the retarded Green
function as shown in Fig.~\ref{fig:figure6}(a) where the lines
$\frac{\omega^2}{\Omega_0^2}$ and $1+\kappa\chi'(\mathbf{Q},\omega)$ has a
single intersection at $\omega=0$. We set the CDW transition temperature $T_{\rm
  CDW}=0.01t$, and different interaction parameters $u$ and $\kappa$ are
selected. The corresponding $1+\kappa\chi'(\mathbf{Q},\omega)$ lines are also plotted in
Fig.~\ref{fig:figure6}(a). All cases in Fig.~\ref{fig:figure6}(a) have no extra
high energy pole beyond the zero frequency. As expected, the dynamic structure
factor exhibits the phonon softening as shown in Fig.~\ref{fig:figure6}(b)-(f),
consistent with the experimental observations in $R$Te$_3$~\cite{Ru2008,Hoesch2009,Maschek2018}.

\emph{Discussions and conclusions. --}
In this work, we mainly study the lattice dynamics for the CDW system evolving
from high temperatures to transition temperature. We present the central-peak
and soft-phonon scenarios for the lattice dynamics that is coupled to different
underlying electronic structures. Both scenarios have a zero-frequency peak
indicating the lattice distortion formation by condensing the collective mode of
the charge-density fluctuation and phonon dynamics. Below the transition
temperature, the zero-frequency mode might be split to a gapless phason mode and
a gapped amplitude one. As we do not perform the self-consistent calculation
below the transition temperature,
such an effect is not included in the present theory.

In conclusion, we consider a simple phenomenological model to study the lattice dynamics in the charge-density-wave metal at
a van-Hove-singularity filling. A VHS-enhanced low-frequency charge-density susceptibility and a
modest electron-electron interaction categorize the CDW-phonon dynamics into the
central-peak scenario which has a central zero-frequency peak and a hardening
CDW-associated phonon mode. Our simple model may have the potential implication
for the absence of the phonon softening in the kagome metal $A$V$_3$Sb$_3$
during the CDW transition. 


\begin{acknowledgments}{This work is supported by National Key Projects for
    Research and Development of China (Grant No.~2021YFA1400400), 
    National Natural Science Foundation of China (NSFC) (Grant No. 11774143), the Guangdong Innovative and Entrepreneurial Research Team Program (Grants No.~2017ZT07C062), Shenzhen Key Laboratory of Advanced Quantum Functional Materials 
   and Devices (No.~ZDSYS20190902092905285), Guangdong Basic and Applied Basic
   Research Foundation (No.~2020B1515120100).}  
 \end{acknowledgments}

\bibliography{VSC}
\end{document}